\documentstyle[preprint,aps]{revtex}
\tighten
\begin{document}
\draft

\title{\bf ANTIBOUND AND RESONANT STATES IN HALO NUCLEI}

\author  {\rm R. Id Betan $^{a,b)}$,
              R. J. Liotta $^{a)}$,
              N. Sandulescu $^{a,c)}$,
              T. Vertse $^{a,d}$}
\bigskip

\address {\rm
  $^{a)}$~  Royal Institute of Technology, AlbaNova University Center,
 SE-10691, Stockholm, Sweden \\
  $^{b)}$~  Departamento de Fisica, FCEIA, UNR,
 Avenida Pellegrini 250, 2000 Rosario, Argentina\\
  $^{c)}$~ Institute of Physics and Nuclear Engineering, 
P.O.Box MG-6, Bucharest-Magurele, Romania\\
  $^{d)}$~ Institute of Nuclear Research of the 
Hungarian  Academy of Sciences,
H-4001 Debrecen, Pf. 51, Hungary}

\maketitle
\begin{abstract}

An unified shell model scheme to evaluate simultaneously the 
contribution of antibound states and Gamow resonances to the nuclear halos
is presented.
The calculations, performed in the complex energy plane, are applied 
to the case of $^{11}$Li. It is found that $^{11}$Li may develop a 
resonant state excitation near the breakup threshold.

\end{abstract}


It is by now well established that halos in 
nuclei are produced by particles moving in single-particle states which
extend far in space. The setting for this to occur
requires neutron configurations consisting mainly 
of unbound $s$- and $p$-waves since in this case no barrier
large enough to trap the nucleons inside the nuclear
core will be present. It is the pairing interaction 
that holds the escaping neutrons together, granting them
their Borromean character \cite{zhu}.

Stated in this simplified fashion the explanation
of halos seems perhaps obvious. However it took many years
of intense work, experimental as well as theoretical, to
reach the understanding of halos. It is not our intention
here to quote the vast amount of papers written on this 
subject. A recent review, including abundant references, can
be found in Ref. \cite{suz}. 
For our purpose it is enough
to point out that for the description of $^{11}$Li, which is the 
typical halo nucleus, the two relevant single-particle states,
as specified by the experimental spectrum of $^{10}$Li \cite{suz,gar}, 
consist of a  low- lying $s_{1/2}$ state, which probably 
is an antibound (or virtual) state at about -50 keV , and a
$p_{1/2}$-resonance at about 540 keV. A second $p$ resonance at around 
0.250 MeV may also exist \cite{bohlen}. The two-body correlations
induce a bound ground state in $^{11}$Li at about -0.295 MeV.
Notice that we are here using the shell-model language, where the core 
($^{9}$Li) is considered as inert, the single-particle states
are given by $^{10}$Li and the two-body nucleus is $^{11}$Li \cite{be}.

The existence of a very low-lying virtual $s$-state in $^{10}$Li has  
important consequences for the correlations developed in $^{11}$Li \cite{tz}.
As discussed below, an antibound state close to the continuum threshold 
enhances the localisation of the low-lying scattering states.
Therefore the $s$-wave content of the ground state of $^{11}$Li is also 
increased, reaching the corresponding (large) experimental value.  
Moreover,  the antibound state in $^{10}$Li can affect
the excited spectrum of $^{11}$Li as well as the ground state.
Since the dineutron system has also a virtual state close to threshold, in 
each of the two-body subsystems of $^{11}$Li there is an antibound state 
near zero energy. Thus $^{11}$Li fulfils the 
Efimov conditions \cite{efimov} and consequently it may form low-lying 
excited states close to the breakup threshold. The posibility of 
such excitations was suggested in Ref. \cite{zhukov94} in connection to 
the momentum distribution of the fragments of $^{11}$Li.

In all the studies done until now the role of the antibound 
$s$-state in $^{11}$Li 
was discussed only indirectly, i. e. through 
the associated scattering length. 
One of the aims of this letter is to present for the first time a formalism
in which antibound states are treated as ordinary single-particle states.
That is, the antibound states will form part of the single-particle
representation in the same fashion as bound states do in 
standard shell model calculations. 
For this we will use an extended version of the shell model on the 
complex energy  plane (CXSM) \cite{rsm,cxsm}. 
Within this formalism we will investigate the existence of 
low-lying excited resonant states of Efimov type, for which there are no 
accurate calculations yet. 

The CXSM is a powerful method to analyse the influence of the
single-particle resonances upon the two-particle correlated states
as well as to understand the developing of bound states by
the two-body force acting upon states embedded in the continuum.
 In its initial form \cite{rsm,cxsm} the representation corresponding 
to the CXSM consists 
of Gamow resonances, bound states and complex scattering states.
These, together with the antibound states, are poles 
of the one-particle Green function and, therefore, we 
will refer to them as "poles".

The space spanned by this basis (Berggren space) \cite{lio} is 
determined by a  metric (Berggren metric) which is non-Hermitian.
Since on the real energy axis the Berggren space coincides with the 
Hilbert space, any physical quantity evaluated within the CXSM coincides
with the corresponding one evaluated within the standard shell model. 
In this sense one can assert that the CXSM is a generalization 
of the shell-model and, as in this model, the many-body energies and 
corresponding wavefunctions can readily be evaluated and analysed.

The CXSM can be easily extended to take into account
the antibound states. The corresponding single-particle representation
contains the antibound states on the same footing as the other discrete
elements, i. e. bound states and Gamow resonances.  
We will show how to construct this representation 
for the case of $^{11}$Li. 

In the first step of the calculation one evaluates the single-particle
states of the unbound nucleus $^{10}$Li. As in Ref. \cite{ebh}, for the 
central field we choose a Woods-Saxon potential with different depths
for even and odd orbital angular momenta $l$.
The single-particle states as well as the scattering waves
will be evaluated by using the 
high precision piecewise perturbation
method \protect\cite{tv2}.
Since for the resonant state $0p_{1/2}$ the experimental data are uncertain
we will perform two calculations corresponding to the energies 200 keV and 
500 keV. At the same time, this will allow us to assess the influence 
of the $p_{1/2}$ resonance upon the halo formation. 
For clarity of presentation we will start studying the 200 keV case
and afterwards the 500 keV case will be presented. At the end we will
compare the results and discuss the similarities and differences
between the two cases. 

For the 200 keV case we use the Woods-Saxon potential given by 
$a$ = 0.67 fm, $r_0$ = 1.27 fm, 
$V_0$ = 50 (36.9) MeV and $V_{so}$ = 16.5 (12.624) MeV for $l$ even (odd).
With these
parameters we found the single-particle bound states 
$0s_{1/2}$ at -23.278 MeV and $0p_{3/2}$ at -2.589 MeV forming 
the $^{9}$Li core. The valence poles are 
the low lying resonances $0p_{1/2}$ at 
(0.195,-0.047) MeV and  $0d_{5/2}$ at (2.731,-0.545) MeV and the
wide resonance $0d_{3/2}$ at (6.458,-5.003) MeV. 
Besides, the state $1s_{1/2}$ appears as an antiboud state\footnote{
The principal quantum number $n$ labelling the single-particle states
indicates that the corresponding wave functions are localized in a region
inside the nucleus and that its real part has in that region $n$ nodes, 
excluding the origin.} at -0.050 MeV. We thus reproduce the experimental 
single-particle
energies giving from the very beginning
unequivocal endorsement to the 
low lying $s$-state  as due to an antibound state.

We also found other resonances at high energies. However
we include in the basis single-particle
states lying up to 10 MeV of excitation energy only. We found that 
expanding the basis from this limit does
not produce any effect upon the calculation up to the six digits of 
precision that we require. 

The non-resonant continuum in the CXSM representation is given
by the points in the contour embracing the resonances \cite{lio}. 
One can use different contours for different partial waves,
but it is important to keep in mind that the corresponding Berggren 
space and therefore the calculated quantities on the complex energy
plane will depend on the contours. Only on the real energy axis the
calculated quantities do
not depend upon the shapes of the contours. 
This property will allow us to check the precision of our calculation.
That is, the calculation of physical quantities by using any contour
should coincide with those evaluated on the real energy axis, i. e. 
by means of the continuum shell-model \cite{ebh}.
Let us stress again that differences between 
the CXSM and the standard shell-model  
appear only when evaluating resonances \cite{cxsm}.

In the calculations presented below we will use the contours shown 
in Fig. \ref{spc}. One sees in this figure that for the
partial waves containing a Gamow resonance we take the conventional
CXSM contour defined by the vertices B \cite{rsm,cxsm} (see also 
Ref. \cite{mic}).
Instead, for the antibound state we introduce the new type of countour
defined by the vertices A. The general requirements which such 
contours should fulfil are discussed in Ref. \cite{tor}. With
proper values for the energies corresponding to the vertices
of these contours one can readily 
include in the basis the three resonances and the antibound state
mentioned above. Notice that in our basis no single-particle bound state 
is present.

The next step in the calculation is to adopt a residual interaction.
We will use a separable force and, therefore,
the Hamiltonian matrix reduces 
to a dispersion relation (for details
see Ref. \cite{cxsm}).  
To evaluate the ground state of $^{11}$Li we adjust the strength 
$G$ of the separable interaction to reproduce the 
corresponding energy, i. e. -295 keV. We thus obtained 
G = 0.00194 MeV. 

With the mean field and the two-body interaction thus established
we evaluated the ground state wave function. First we performed 
the calculations by choosing 
the real energy as a contour. In this case the wave function is 
spread over many  components. The largest of these components
correspond to configurations $p_{1/2}\otimes p_{1/2}$ lying
close to 400 keV (i. e. about
twice the energy of the $0p_{1/2}$ resonance)  and 
$s_{1/2}\otimes s_{1/2}$ lying close to threshold 
(i. e. close to twice the energy of the antibound state). 
The wave function consists of 44 \% $s$-states, 48 \% $p$-states
and 8 \% $d$-states, as expected \cite{suz,gar}.

A remarkable feature of the calculation is that the antibound 
state exerts such a strong effect upon the two-particle wavefunction. 
From a CXSM point of view this is because the
energy corresponding to the configuration $(1s_{1/2})^2$ is 
very close (in the complex energy plane) to the two-particle energy.
But it can also be understood from a continuum shell-model
point of view. That is, the radial scattering wave function 
with energies $E=\hbar^2 k^2/2\mu$ ($k$ real
and positive) close to a bound or antibound state with energy
$E_0$ ($k_0=\pm i|k_0|$, respectively)  
lying near threshold can be written as \cite{mig},
\begin{equation}\label{eq:mig}
{\cal R}_l(kr) \approx \sqrt{2k|k_0|\over{k^2+|k_0|^2}} 
{\cal R}_l(|k_0|r)
\end{equation}
which shows that close to threshold the radial dependence  of
the scattering wave functions is independent of the
energy, except the square root factor.
This factor is maximum for $k=|k_0|$.
Therefore, the matrix elements of the two-body
interaction are large within a relative 
large interval close to threshold. This induces large components 
of the two-body wave function in that energy region.
Notice that it does not matter whether the 
S-matrix pole $E_0$ corresponds to a bound or to an antibound
state. The effect is exactly the same.

This feature can be seen in Fig. \ref{anti}, where we give the 
localization of the scattering states, $L(E)$, defined as 
\begin{equation}\label{eq:loc}
L(E)=\int_0^{1.2R_N} {\cal R}^2_{l=0}(kr) r^2 dr
\end{equation}
where $R_N$ = 2.6 fm  is the core nuclear radius. 

One sees in Fig. \ref{anti} a strong increase of the 
localization as the energies of the poles approaches threshold.
This propery is also responsible for a similar increase in the elastic 
cross section, as seen in Figs. 12 and 13 of Ref. \cite{amb}. 
Therefore one expects large elastic cross sections  
at low energies in halo nuclei, particularly in $^{11}$Li. 

So far  we have shown the advantages of the CXSM to evaluate the effect
of the antibound state on already known properties of the ground state
of the halo nucleus $^{11}$Li. However, the transparency of the method 
becomes essential in the search for other physically meaningful
two-particle states in the continuum. In what follows we discuss
the problem of low-lying $0^+$ excitations and their influence upon 
the neutron halo.

The possibility of low-lying excitations in $^{11}$Li, with energies
below the first experimentaly measured excitation at 1.2 MeV, was
considered in Ref.\cite{zhukov94}. In this reference it is argued
that a low-lying narrow resonance at around $0.2-0.4$ MeV
would be consistent with the momentum distribution of the
$^{11}$Li fragments. To our knowleadge the existence of such a low-lying 
resonance in $^{11}$Li was not investigated in microscopical 
calculations so far. 

Within the CXSM two-particle resonances are easy to calculate
since they appears as a result of the diagonalization of the Hamiltonian 
(which in our case reduces to the solution of the dispersion relation) 
in the complex energy plane. Thus we found that the first excited state 
(i. e. the state $0^+_2$) appears at the complex 
energy (0.202,-0.137) MeV. The corresponding wave function consists of 
nearly 100 \% $p$-states, with a small admixture of $s$-states.

It is interesting to analyse how this state is built up by the
two-body interaction starting from the zeroth-order configuration 
$(0p_{1/2})^2$. For this we increased the 
interaction gradually starting from G=0, as seen in Fig. \ref{li200}.
As the attractive interaction increases the resonance becomes 
narrower and approaches threshold,
as expected from perturbation theory. 
However, a point is reached where
continuum configurations become important and the resonance widens.
This happens at G=0.0005 MeV in the figure. Up to this point the resonance
is a purely  $(0p_{1/2})^2$ state and, therefore, it is localized
inside the nucleus. That is, it is a physically meaningful resonance.
But from here on other configurations
become important. These configurations 
are overwhelmingly those where one 
of the particles moves in the continuum and the other 
in the resonance $0p_{1/2}$. More
specifically, in the two-particle resonace that we are studying 
the most important states in the 
continuum are those corresponding to $p_{1/2}$ waves with energies 
corresponding to the points on the segment $B_1-B_2$ in Fig. \ref{spc}.
At $G=G_0$=0.00194 MeV, corresponding to the G-value fitting the ground
state, there is a strong mixing with the continuum configurations.
Increasing $G$ farther the configuration $(0p_{1/2})^2$ looses its 
importance, the resonance is split in a number of pieces and 
eventually dissolves into the continuum. However, since these
configurations virtually include only $p$-waves the wavefunctions 
still consist of only $p$-states.

The analysis that we have done so far is based upon the assumption
that the $p$ resonance in $^{10}$Li is located at 200 keV.
To see the influence of this resonance on the structure of the halo
we will now analyse the ground and the excited states of $^{11}$Li by
using the 500 keV case.
For this we adopted the Woods-Saxon depth $V_0$ = 35.366 MeV 
for odd $l$-values, keeping all other parameters
as before. We thus obtained the energy (0.470,-0.197) MeV 
for the state $0p_{1/2}$.  The state $0p_{3/2}$, belonging
to the core, is found now at -2.016 MeV. The other odd l-value
poles lie beyond the range of energies included
here. But, nevertheless, we have checked that they do not affect the 
results.

We kept the two-particle interaction used in the previous case, except
that the strength  necessary to adjust the energy of $^{11}$Li(gs)
is now G = 0.00694 MeV. 

As before, we found that on the real energy axis the ground state
wave function is spread in many components. The largest of them lie 
close to threshold for the configurations $s_{1/2}\otimes s_{1/2}$
and around 1 MeV for the $p_{1/2}\otimes p_{1/2}$ configurations.
The wave function consists of 49 \% $s$-states, 
39 \% $p$-states and 12 \% $d$-states, which is also within the 
range of accepted values \cite{suz,gar}.

Since the position of the $p_{1/2}$ pole seems likely to correspond to the 
present 500 keV case \cite{gar}, we will analyse here 
the effects of the antibound and
the Gamow poles upon the ground state of $^{11}$Li by using 
the contours of Fig. \ref{spc}. 
We therefore present in Table \ref{cgs} the contribution of 
different configurations to that ground state.
The corresponding complex amplitudes depend on the chosen contours 
and have no direct physical meaning. But the total content of a given
partial wave in the bound ground state wave function, which is a physical
quantity, does not depend upon the chosen contour. From Table \ref{cgs} we
can see that for the $p$ and $d$ waves the configurations are built 
mainly on the corresponding Gamow resonances. The situation is different
for the $s$-wave since apart from the configurations built upon the antibound
state there is also an important contribution 
coming from the complex scattering
states. This contribution is given mainly by those $s$ scattering states 
located on the segments $(0,0)-A_1$ and $A_1-A_2$ of Fig. \ref{spc}, which
are the closest to the antibound state. 

Up to this point there is not much difference between the 200 and
the 500 keV cases, which may explain why various studies of 
the halo structure of $^{11}$Li(gs) with the common
feature of having low-lying s- and p-states, provide similar results
\cite{gar}. This is because the wave function of $^{11}$Li(gs) 
is mainly controlled by low-spin single-particle states lying
close to the continuum threshold.
The exact positions of the resonances do not influence the wave function 
very much. Even the two-particle interaction (if reasonable)
is not so important in the evaluation of the properties 
of $^{11}$Li(gs) since
this state is a Cooper pair  \cite{coo} and as such is mainly induced by 
the Pauli principle acting upon the valence particles and 
those in the core. 
However, the two-body interaction as well as the position of the
single-particle poles may have a fundamental importance to 
determine the physically meaningful excited states. The states
arising from the particles moving in the continuum  
are not localized inside the nucleus and, therefore, will
be weakly affected or not affected at all by the interaction.
This can be seen in Fig. \ref{li500}, where we present the evolution
of the state $0^+_2$ as a function of the strength G.
The $0p_{1/2}$
resonance is now wider and higher in energy than before. As a result,
the point corresponding to G=0 in the figure is closer to points 
coming from the continuum contour. Yet, these continuum states
do not seem to affect the resonance as G increases. That is,
the behaviour of the resonance as G is varied is very similar
to that in Fig. \ref{li200} as well as to resonances in non-halo nuclei 
\cite{cxsm}. The reason for this is that
those points, label "s-states" in the figure, correspond to configurations
of the type $cs_{1/2} 1s_{1/2}$, where "c" labels points in the segments
$(0,0)-A_1$ and $A_1-A_2$ of Fig. \ref{spc}. The overlap 
between these configurations
and the mainly $(0p_{1/2})^2$ configuration of the
resonance is small. Only at large values
of G (above G=0.004 MeV in the figure), the resonance starts to feel 
the presence of the continuum states. 
The remarkable feature of the
figure is the sudden turning down of the curve corresponding to the
physical resonance at G=0.005 MeV. As 
G increases in this region the continuum plays a mounting role.
As in the 200 keV case above, the most important of the continuum
configurations are those in which one particle moves in the continuum
and the other in the $0p_{1/2}$ resonance. There could be many comparatively
large configurations of these type and it would not be useful to give
all of them. More instructive is to show their contribution to the 
normalization of the wave function in this case where, in contrast 
to the ground state case of Table \ref{cgs}, the zeroth order energies 
are not very close to the energy of the state $0^+_2$. We thus define $S(l)$ as 
the sum of the squares of 
the amplitudes corresponding to configurations where at least one of
the two particles moves in continuum states. 
On the real energy axis the sum of $S(l=1)$ and $X^2((0p_{1/2})^2)$,
where $X$ is the wave function amplitude, is the probability of 
the p-content of the wave function. This is a 
quantity that we evaluated above for the ground state.
The dependence of these quantities upon G corresponding to the state $0^+_2$ 
is shown 
in Table \ref{liwf}. Since we are studying states with 
complex energies the numbers $S$ as well as $X^2$
are complex in this Table. Moreover, their absolute values could 
be larger than 1
although the sum of all possible l-contributions is normalized to (1,0). This
is a good example of the non-Hermitian character of the Berggren metric.

One sees in this Table that the two-particle resonance starts to mix 
with the continuum at G= 0.005 MeV and at G=G$_0$ it is composed 
mainly of continuum configurations. Therefore, at this point it has
already lost its localization features. It has become a part of the
continuum background.

We are now in a position to recognize other systems where halos may
be present. We thus looked for nuclei which may be considered
shell-model cores lying on the neutron drip line with  
low-lying single-particle resonances carrying low-spin.
Following the trend of single-particle states in the
relativistic mean field calculations we found that 
Z=20, N=50 may be such a core. In order to simulate the order
of the single particle states given by the relativistic calculations
we used a Wood-Saxon potential defined by
$a$ = 0.67 fm, $r_0$ = 1.27 fm,  $V_0$ = 39 MeV and $V_{so}$ = 22 MeV. 
With this potential the antibound $2s_{1/2}$  state (note that $n$=2)
appears again 
at -0.050 MeV. But now the next valence shells are $1d_{5/2}$ 
at (0.469,-0.048) MeV,
$1d_{3/2}$ at (2.080,-1.525) MeV, $0g_{7/2}$ at (6.739,-0.738)
MeV and $0h_{11/2}$ at (5.344,-0.102) MeV. The states in the
core are ordered as usual. As expected, the highest of these is the 
state $0g_{9/2}$, lying at -2.276 MeV. Using the same separable 
interaction as before and assuming again that the ground state 
of $^{72}$Ca lies at -295 keV, we obtained for the strength of the 
interaction the value $G_0$=0.00174 MeV. Close to this $G_0$ value
we found also a low-lying two-particle resonance with the energy of 
about (0.550, -0.350) MeV. The behaviour of this $0^+_2$ 
resonance as a function of $G$  is very similar to the
500 keV case of  Fig. \ref{li500}. The discussion performed there
is also valid here and, therefore, we will not analysed this rather
academic case farther. But it is important to point out that in this
and the other
cases presented here, we have been careful to choose contours that leave the 
region around the two-particle resonances in the complex energy plane
free of continuum configurations. We thus established an "allowed" region
\cite{cxsm}. Otherwise the two-particle resonance would be embedded 
in a see of continuum states, making the calculations difficult and
the evaluated quantities unreliable.

In conclusion, we have presented in this letter a formalism that 
treats the antibound states exactly in the same fashion as bound 
states and Gamow resonances in the framework of a shell model 
scheme in the complex energy plane. We found, as expected, that 
antibound states lying close to the continuum threshold are of a fundamental 
importance to build up the halo. But we also found that an excited low-lying 
two-particle resonance may exist in halo nuclei. For the case of $^{11}$Li
this low-lying resonant appears in the energy range of $0.2-0.5$ MeV. 
This energy range is the same as the one suggested 
in Ref. \cite{zhukov94} in connection to the momentum distribution of 
$^{11}$Li fragments. However, the calculations exibit a very drastic change 
in the structure of the resonant excited state when the strength of the force 
is approaching the value used for the determination of the ground state. 
Do to this it is rather difficult to conclude whether this state is a physical 
resonance or not. Further  efforts directed to the measuring of the eventual 
low-lying resonance excitations in $^{11}$Li would be essential in order to 
settle this issue.

\acknowledgments

This work has been supported by FOMEC and Fundaci\'on
Antorcha (Argentina), by
the Hungarian OTKA fund Nos. T37991 and T29003 and by
the Swedish Foundation for International Cooperation
in Research and Higher Education (STINT).

\begin{figure}
\caption{In the CXSM the contour is replaced by integration
points \protect\cite{lio}. The triangles indicate the 
points belonging to the $l=0$ contour used to incorporate the antibound
state in the Berggren basis. The circles indicate the points
corresponding to the contour embracing the resonances with $l\not= 0$.}
\label{spc}
\end{figure} 

\begin{figure}
\caption{Localization $L(E)$, Eq. \protect\ref{eq:loc}, in the presence
of low-lying antibound (a)
and bound (b) $s$-states. The numbers labeling the curves are the energies 
of the poles in MeV.}
\label{anti}
\end{figure} 

\begin{figure}
\caption{Evolution of the energies of the two-particle resonance 
$^{11}Li(0^+_2)$ as a function of G (in MeV) for the 200 keV case.
The numbers labeling the open circles are the values of $G\times 
10^4$ MeV. 
The strength adjusted to obtain the ground state is $G_0$ = 19.4
$\times 10^{-4}$ MeV.}
\label{li200}
\end{figure} 

\begin{figure}
\caption{As Fig. \protect \ref{li200} for the 500 keV case except
that the numbers are the values of $G\times 10^3$ MeV and 
$G_0$ = 6.94 $\times 10^{-3}$ MeV}
\label{li500}
\end{figure} 

\begin{table}
\caption{ The contribution of  partial wave configurations
 $(s_{1/2})^2$, $(p_{1/2})^2$ and $(d_{5/2})^2$ to the ground state
wave function calculated in the complex energy plane. For each partial wave are
given the square amplitude of the  pole-pole term and the sum of the square 
amplitudes corresponding to the pole-scattering and scaterring-scattering
terms. The total contribution of each partial wave is given in the last line.
\label{cgs}}
\begin{tabular}{cccccccc}
            &  $(s_{1/2})^2$      & $(p_{1/2})^2$   & $(d_{5/2})^2$ \\
pole-pole   & ( 12.936, -0.039 )  & (0.642, -0.204) & (0.127, 0.011)\\
pole-scat.  & ( -29.365, 0.079 )  & (-0.279, 0.221) & (-0.031, -0.018)\\
scat.-scat. & ( 16.921, -0.040 ) & (0.022, -0.017) & (-0.002, 0.007) \\
total       & ( 0.492, 0.0)      &  (0.385, 0.0)   &  (0.094, 0.0)
\end{tabular}
\end{table}

\begin{table}
\caption{Square of the wave function amplitude $X((nlj)^2)$,
where $(nlj)$ indicates the quantum number of the single-particle resonance,
and the sum S(l) corresponding to the physical resonance of Fig.
\protect\ref{li500}. The values of G are in MeV and $G_0$ = 0.00694.
\label{liwf}}
\begin{tabular}{cccccccc}
G&$X^2((1s_{1/2})^2)$&S(0)&$X^2((0p_{1/2})^2)$&S(1)\\
0.001&(0.00,0.00)&(-0.00,0.00)&(1.00,0.00)&(0.00,-0.00)\\
0.003&(0.05,0.02)&(-0.10,0.10)&(1.06,0.06)&(0.00,-0.05)\\
0.005&(-0.01,0.11)&(-0.25,-0.12)&(0.59,0.30)&(0.47,-0.19)\\
$G_0$&(-0.04,0.02)&(-0.00,-0.11)&(0.23,0.18)&(0.74,-0.12)\\
0.008&(-0.03,0.01)&(0.01,-0.08)&(0.18,0.14)&(0.79,-0.10)\\
\end{tabular}
\end{table}

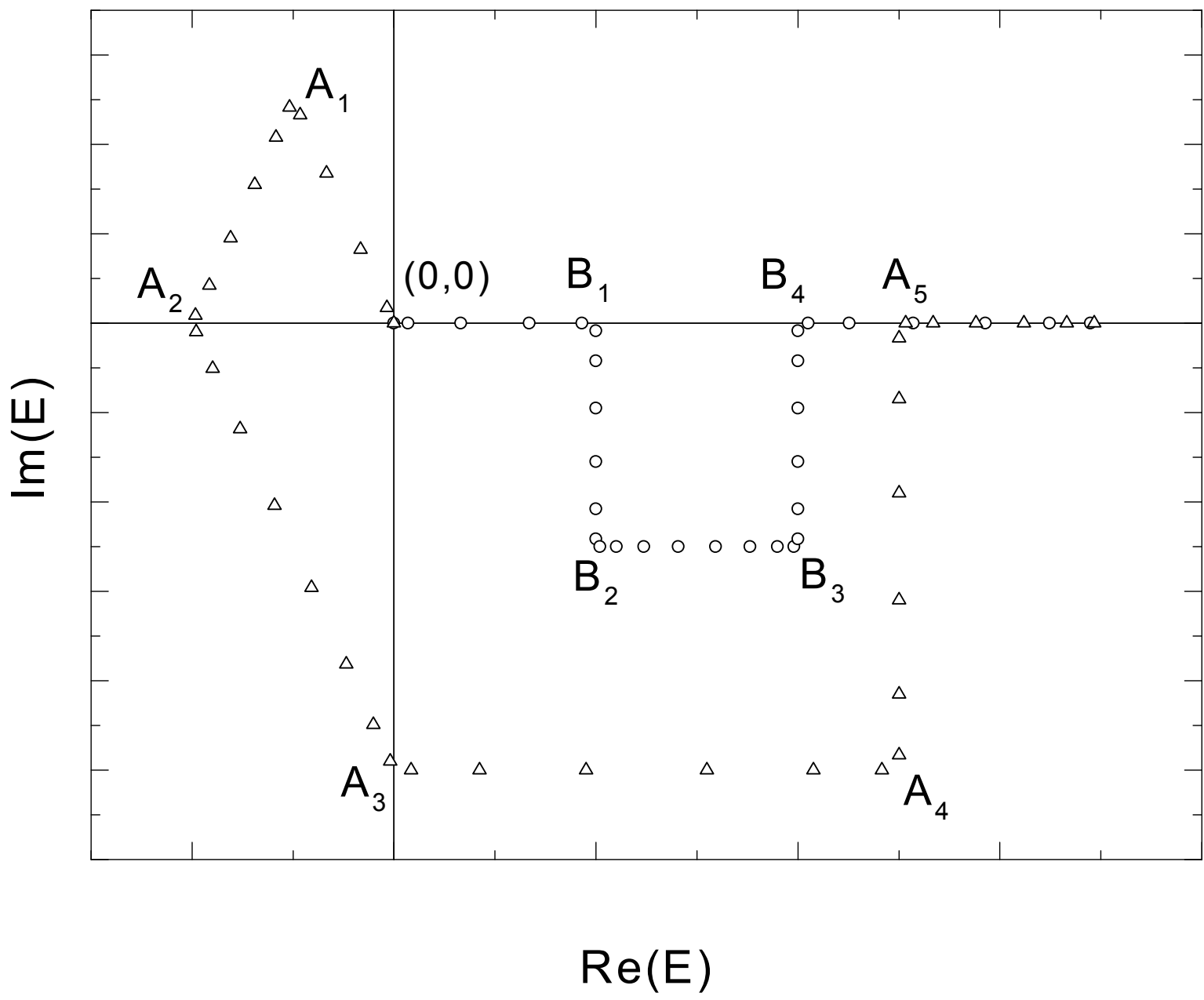
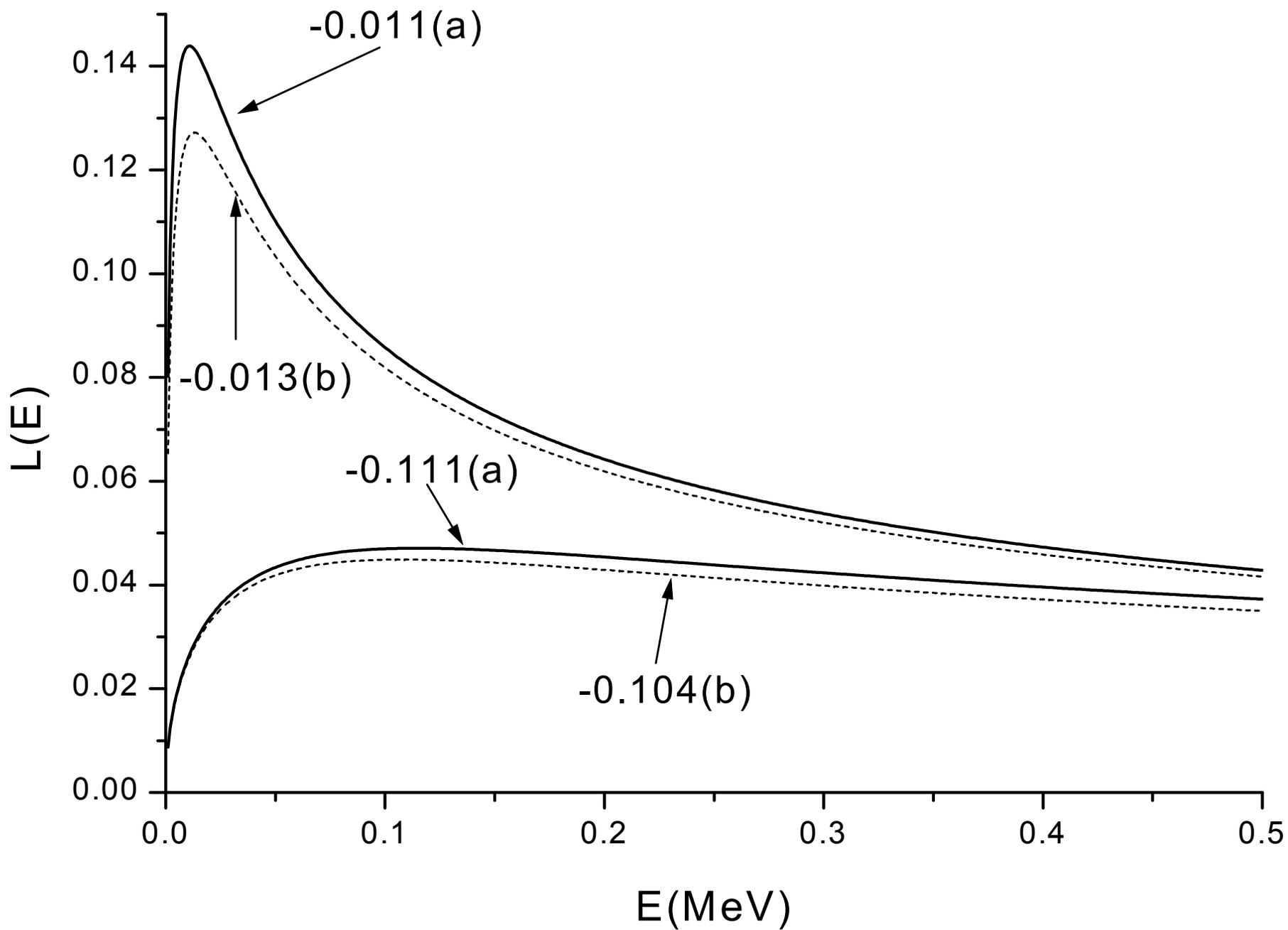
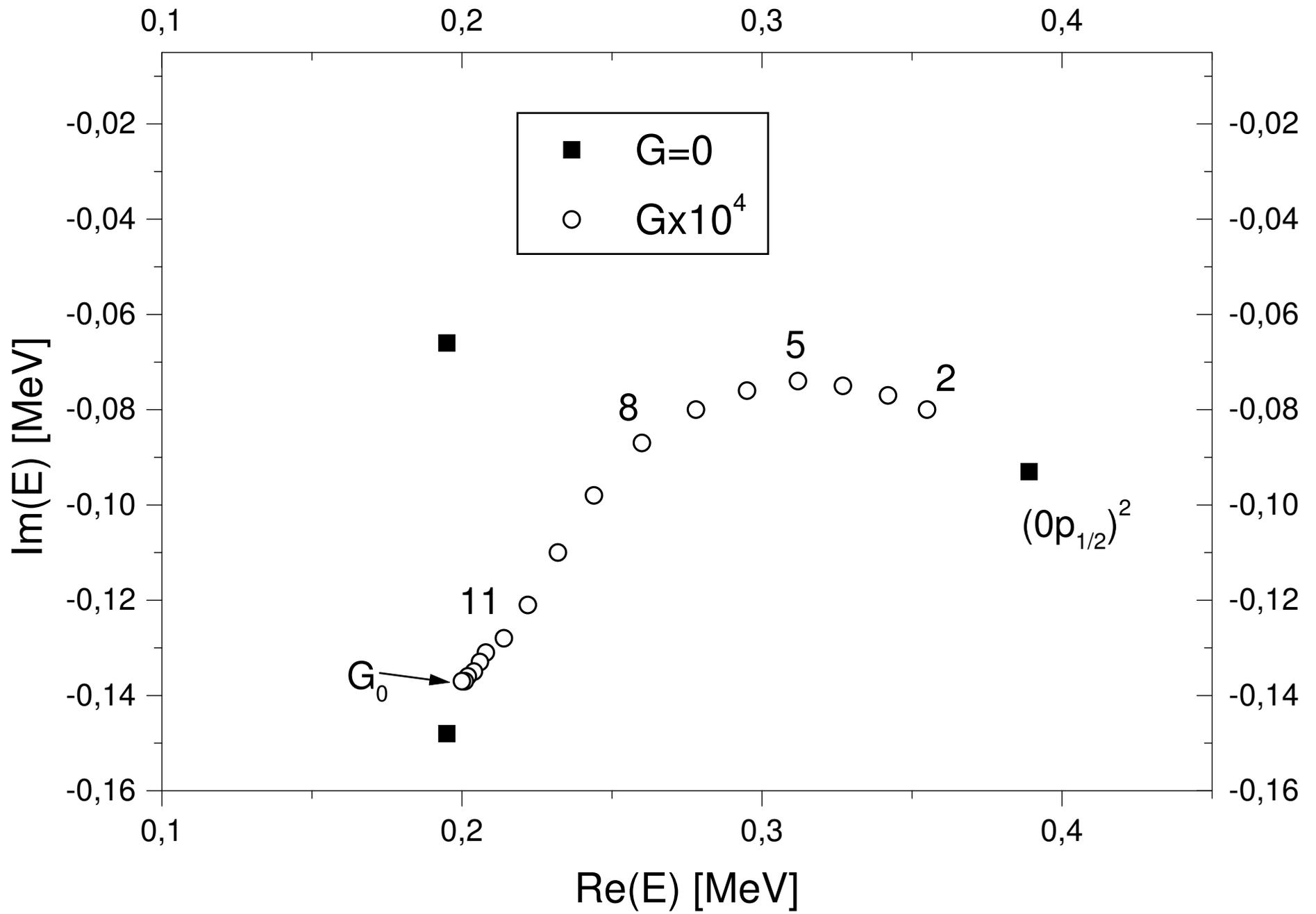
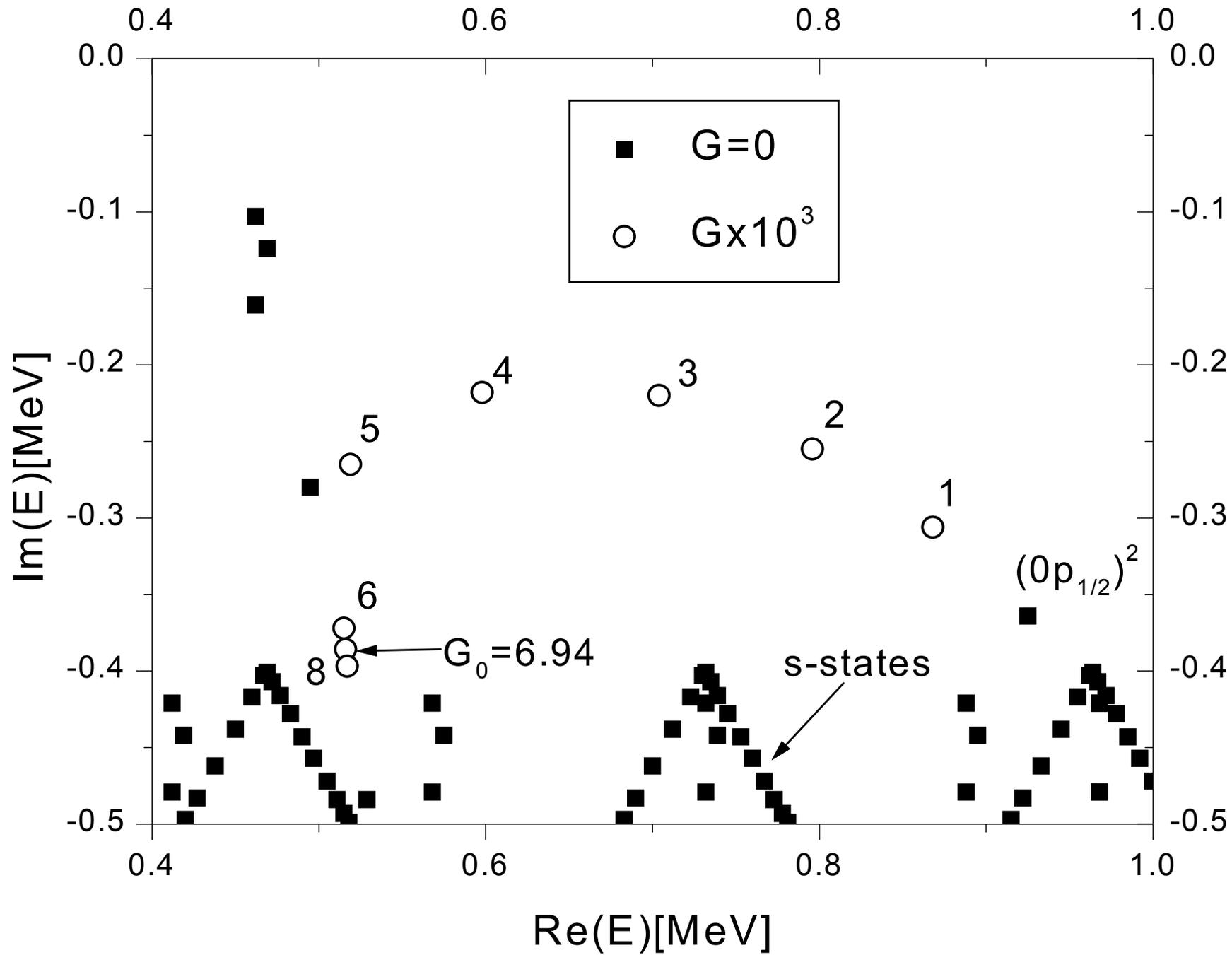
\end{document}